\documentclass[aps]{revtex4}
\usepackage{graphicx}

\begin{document}

\title{Binary fluid demixing: The crossover region}

\author{ Ignacio Pagonabarraga\\
Departament de F\'{\i}sica Fonamental, Universitat de Barcelona, \\
Av. Diagonal 647, 08028-Barcelona, Spain\\
 Alexander J. Wagner and M.E. Cates\\
Department of Physics and Astronomy, University of Edinburgh,\\
                            JCMB Kings Buildings, Mayfield Road,
                            Edinburgh EH9 3JZ, U.K.}

\begin{abstract}
By performing lattice Boltzmann simulations of a binary mixture, we
scrutinize the dynamical scaling hypothesis for the spinodal
decomposition of binary mixtures for the crossover region, {\em i.e.}
the region of parameters in the growth curve where neither inertia nor
viscous forces dominate the coarsening process. Our results give no
evidence for a breakdown of scaling in this region, as might arise if
the process were limited by molecular scale physics at the point of
fluid pinch-off between domains. A careful data analysis allows us to
refine previous estimates on the width of the crossover region which
is somewhat narrower than previously reported.

\end{abstract}
\maketitle

\section{Introduction}

Spinodal decomposition occurs when a binary mixture, which forms a
homogeneous phase at high temperature, undergoes spontaneous demixing
after imposing a sudden quench to below the spinodal temperature of
the mixture. Following the quench, after a short period of
interdiffusion, the mixture will form domains of different
composition, separated by sharply defined interfaces. In its late
stage, the local compositions of the fluid domains correspond to those
of the two bulk phases at coexistence, while the interfacial tension
approaches its equilibrium value. The shape and evolution of the
domains will depend on the initial composition of the mixture, as well
as on the different physical parameters characterizing the fluids. For
the case of a symmetric mixture, in which the amounts and properties
of the two immiscible species are the same, then starting from a
randomly mixed initial state, bicontinuous structures develop.

As emphasized by Siggia\cite{siggia} and Furukawa\cite{furukawa}, the
physics of spinodal decomposition involves capillary forces, viscous
forces and fluid inertia. Local interfacial curvature generates stress
which drives fluid motion; this will propagate by viscous force or
inertial motion (or both) depending on the parameters of the fluid. If
we assume no other physics determines the spinodal process, the
parameters determining the fluid behavior are then the interfacial
tension, $\sigma$, fluid mass density, $\rho$, and viscosity $\eta$
(for a deep quench, diffusion does not contribute significantly beyond
a short initial transient). Any degree of asymmetry in the mixture's
composition or in the dynamical properties of the fluids will provide
additional control parameters.

For the completely symmetric case, from the three relevant
fluid parameters only one length, $L_0$, and one time, $T_0$, can be
constructed: $L_0=\eta^2/(\rho \sigma)$ and $T_0=\eta^3/(\rho
\sigma^3)$. The absence of other physical mechanisms controlling the
spinodal decomposition in the late-stage domain growth leads to the
{\em dynamical scaling} hypothesis\cite{siggia,furukawa}. According to
it, if we express the domain size, $L(T)$ as a function of time $T$ in
reduced units using the characteristic length and time scales, the
corresponding dimensionless length $l\equiv L/L_0$, expressed in terms
of the dimensionless time $t\equiv T/T_0$, will be a universal scaling
function 
\begin{equation} 
l = l(t) 
\end{equation} 
the same for all
fully symmetric, binary incompressible fluid mixtures. The use of
scaling concepts has been of extreme help in understanding and
rationalizing the physics underlying complex phenomena. 
 In the present problem, the dynamical scaling states that all fluids
will have a unique domain growth rate governed only by the value of
$L/L_0$, which  controls the dominance of viscous or inertial fluid
transport. Therefore, this scaling expression provides a systematic
and simple way to analyze the spinodal decomposition of any fluid, in
terms of its basic physical mechanisms.

Furukawa (following Siggia) \cite{furukawa,siggia} analyzed the
limiting behavior of the proposed universal function $l(t)$. For small
enough $t$ fluid inertia is negligible compared to viscous forces,
while for large $t$ the opposite is true\cite{inertial}. Balancing the
dominant terms in the Navier-Stokes equation in each case leads to the
asymptotes 
\begin{equation} 
l(t) \rightarrow\left\{ \begin{array}{ll}
                b t &; \;\;\;\;\; t \ll t^* \\
                c t^{2/3} &;\;\;\;\;\; t \gg t^*
  \end{array}\right.
\end{equation}
where $t^*$ stands for the crossover time, when the viscous and
inertial contributions are comparable. This can be defined more
precisely as the time at which the extrapolated viscous and inertial
asymptotes of the $l(t)$ cross each other on a log-log plot. For
intermediate times around $t^*$, the universal function $l(t)$ will
exhibit a smooth crossover from the viscous to the inertial asymptotic
behavior. It is important to remark that dynamical scaling implies the
universality, not just of the asymptotic power laws for the scaling
function $l(t)$, but of the entire curve, and accordingly of the
amplitudes $b$, $c$ and $t^*$.

Although there is good experimental evidence for the viscous growth
regime\cite{expt1}, the inertial regime is more difficult to study
experimentally\cite{viv2prl,vivjfm}. Therefore, simulation techniques
have so far been the basic tools to analyse the different growth
regimes and validate the scaling hypothesis
\cite{vivjfm,simulations,vivprl,jury}. (The ease with which fully
symmetric immiscible fluids can be created on a computer, compared to
the laboratory, is also a factor.) Such scaling has allowed
classification and comparison of (previously scattered) simulation
results as well as a critical analysis of the derived parameters (such
as $b$), revealing a number of discrepancies in the published data
(mainly related to residual diffusion and/or finite size
effects)\cite{vivjfm}. In our view, the existence of a viscous scaling
regime (with a universal $b$) is now fairly clearly established in
three dimensions. (The case of two dimensions is quite different
\cite{wagner,wagner2D}.) The search for an asymptotic inertial regime
has likewise been extended carefully up to Reynolds numbers of order
300 (with Reynolds number defined as $Re = L \dot{L} \rho/\eta = \dot
l l$). Although there is still a controversy regarding the true final
asymptote at sufficiently high Reynolds number\cite{grant,viv2prl},
previous work\cite{vivjfm,vivprl} does establish the existence of a
well defined asymptotic inertial regime, at least as far as the $l(t)$
curve is concerned, at relatively modest $Re \ge 100$. The attainment
of asymptotic behaviour in this region of $l(t)$ (roughly, $3 \times
10^3 < l < 3\times 10^5$) is less obvious when one examines velocity
statistics\cite{vivjfm} or interfacial dynamics\cite{njp}. However,
further exploration at still higher $Re$ (say, $l\ge 10^6$) remains
out of reach with present simulation methods.

Compared to the status of the viscous and inertial regimes, the
existence of dynamical scaling in the crossover regime remains to be
established. So far it was reported in \cite{vivjfm,vivprl} that the
crossover region spans four decades in dimensionless time, and that
the crossover time $t^*$, although formally of order one, is about
$10^4$. But the lattice Boltzmann data presented in \cite{vivjfm} is
in fact relatively sparse for the crossover region, and does not allow
a direct check of the proposed collapse onto a universal curve.

There are, however, some results found using dissipative particle dynamics
\cite{jury} which lie within the crossover regime as determined in
\cite{vivjfm}. Jury et al \cite{jury} reported incomplete
collapse of this data: each dataset had a similar slope but the curves
did not join up on the $l(t)$ plot. Their data was found to be broadly
compatible with logarithmic deviations from dynamic scaling. Although
the more recent (lattice Boltzmann) results obtained in \cite{vivprl}
suggest that those deviations may instead be due to finite size
effects, the validity of dynamical scaling within the crossover region
is far from completely established.

Jury {\em et al.} noted that the dynamical scaling hypothesis relies
on the assumption that viscous forces and inertia are the {\em only}
physical mechanisms controlling domain growth.  Other physics is,
however, involved too: we have already mentioned diffusion, which can
however be kept negligible, on the domain scale, for deep quenches
(see Ref.\cite{vivjfm} for a discussion of this). Another mechanism
that must be present during the domain growth arises during
topological reconnection or {\em pinch-off} events. These events are
necessary for the coarsening of the bicontinuous structure. They
correspond to the contraction of a fluid neck to zero width in a
finite time (see e.g. Ref.\cite{njp} for visualizations of the
evolution and rupture of necks in the different regimes of spinodal
decomposition).  Unless diffusion intervenes first, the final stage of
such a reconnection event always involves a microscopic, molecular,
length scale. Moreover, recent studies indicate that such reconnection
events follow a distinct scaling of the dynamics\cite{eggers,brenner}
that might, in principle, interfere somehow with the scaling law for the domain
size. It is not clear whether this could happen only in the crossover
region and not in the viscous or inertial asymptotes, but nonetheless,
there is enough uncertainty to warrant a detailed study of the
crossover domain, which we now present.

\section{Model}

Spinodal decomposition with a deep quench is essentially a problem of
deterministic, isothermal fluid motion coupled to moving interfaces,
in which it is important to deal with pinch-off events
consistently. As already mentioned, diffusion is irrelevant, and
thermal fluctuations can also be disregarded at long times. To study
this process we have therefore used the lattice Boltzmann method. This
is a simulation technique in which the interfaces emerge as a result
of the imposed free energy.  This fact has some advantages with
respect to other standard numerical techniques to study Navier-Stokes
equations. The details of the model for the case of a binary mixture
we have used are described elsewhere\cite{jccpc,vivjfm,swift}. The
model simulates a binary mixture with the model free energy
\begin{equation} 
F=\int d{\bf r}
\left\{-\frac{A}{2}\phi^2+\frac{B}{4}\phi^4 + \frac{1}{3}\rho\log\rho +
 \frac{\kappa}{2}|\nabla\phi|^2\right\},
\label{eq:freeen}
\end{equation}
in which $A$, $B$ are parameters that control the phase behavior of
the mixture. When $A < 0$ and $B > 0$ two phases coexist in
equilibrium. In Eq.\ref{eq:freeen}, $\phi$ is the usual order
parameter (the normalized difference in number density of the two
fluid species),and  if
 $A=B$, the order parameter in the coexisting low temperature
phases takes values $\phi=\pm 1$. (This is a matter of convention, not
of physics.) The parameter $\kappa$ is related to the energy cost of
generating a spatial gradient. It determines the value of the
interfacial width, $\xi=5\sqrt{\kappa/2A}$, and the surface tension
$\sigma=\sqrt{8\kappa A^3/9B^2}$. Finally, $\rho$ is the total fluid
density, which remains essentially constant during phase separation.
(This is ensured by working at very low Mach numbers.)

The lattice Boltzmann model for a binary mixture involves two velocity
distribution functions, $f$ and $g$, on a discrete lattice with a
discrete time relaxational dynamics. The properties of the model are
fixed through the equilibrium distributions towards which $f$ and $g$
relax; the zeroth, first and second moments of $f$ determine the
density, the momentum flux, and the stress tensor, while moments of
$g$ determine $\phi$ and the order parameter flux. In this way the
dynamics of the distribution functions is connected to the macroscopic
behavior of the fluid. It has been shown that the Navier-Stokes and
advective-diffusion equations are recovered as the hydrodynamic limit
of the corresponding lattice Boltzmann dynamics\cite{swift}, at least
in the incompressible limit\cite{vivjfm}. The form of the pressure
tensor depends on the free energy model chosen\cite{swift}.

The spontaneous emergence of interfaces as a result of the imposed
free energy, eq.(\ref{eq:freeen}), is an appealing feature of this
method. However, as a result the interfaces are not structureless, but
have finite width $\xi$. In practice, the parameters of the free
energy are chosen such that $\xi$ remains around 3 lattice spacings in
all runs; this minimizes anisotropy effects due to the underlying
lattice. The finite width imposes restrictions on the length scales
that can accurately be studied with this method. In real fluids
(deeply quenched) the interfacial width is a few atomic diameters,
whereas we are often in the regime of $\xi  \gg L_0$ (especially for
the most inertial runs), which is much larger. For spinodal
decomposition this should not matter as long as the dynamical scaling
hypothesis holds. However, if the interfacial width turns out to be
dynamically relevant, as could possibly apply during pinch-off, the
variation between runs of $\xi/L_0$ has to be taken into account.

Since
we are interested in testing the dynamical scaling hypothesis in the
absence of diffusion, the mobility must also be chosen to ensure that,
after a short  initial transient which is always dominated by
diffusion (when the initial domains are formed) diffusion gives an
irrelevant contribution to the growth rate. This can be done using
protocols described in\cite{vivjfm}.

\section{Method and results}
\label{sect:res}

We have performed a number of new lattice Boltzmann simulations on
large ($256^3$) lattices, with parameters chosen to explore the
crossover region, as well as a couple of runs deeper into the viscous and 
inertial regimes to clarify the convergence to the corresponding 
asymptotic regimes. These add significantly more data to those presented
previously in \cite{vivprl,vivjfm}, where the focus was on the
asymptotes rather than the crossover regime. We start with random
initial conditions (effectively infinite temperature), and choose the
parameters in the free energy, eq.(\ref{eq:freeen}), such that $B>0$
and $A=-B$. This is equivalent to performing a sudden deep quench
below the spinodal temperature of the mixture (which corresponds to
$A=0$). We then follow the subsequent evolution of the system by
computing the characteristic domain size $L(T)$, which is then
converted to reduced physical units.

There are several different ways to measure the domain size in a
binary mixture\cite{vivjfm}. Here we choose to employ a characteristic
length scale extracted from the curvature of the interface (see
Ref.\cite{wagner}). The latter is a tensor, defined by
\begin{equation} 
D_{\alpha\beta} = \frac{\sum_{lattice}
\partial_{\alpha}\phi \partial_{\beta}\phi}{\sum_{lattice}\phi^2}
\end{equation} 
where $\alpha$ and $\beta$ refer to the three spatial
coordinates, and $\partial_{\alpha}$ means the spatial derivative with
respect to the $\alpha$ component of the spatial coordinate. The three
eigenvalues of this matrix $\lambda_i$ are then three characteristic
inverse lengths. From them, the domain size, $L$, can be estimated as
\begin{equation} 
L=\frac{3}{\lambda_1+\lambda_2+\lambda_3}
\label{eq:lt}
\end{equation}
Strictly speaking, this is a square of a length; it is related to the 
product of two characteristic lengths. However, the second relevant 
length for the order parameter gradients in the system is the interfacial 
width. This is the same for all the simulations; hence, it contributes 
a constant factor of order unity to eq.(\ref{eq:lt}).
We have monitored the values of the domain size as a function of time
after the quench, $L(T)$. Before seeking data collapse by converting
to reduced physical units, it is very important to select the subset
of reliable data from each simulation run. Early in each run one
always has a period of interdiffusion, followed by a regime where
diffusion and hydrodynamics are both present. Only when the diffusive
contribution to the coarsening rate becomes small can the $L(T)$ data
be expected to scale onto the universal $l(t)$ curve. (We removed the
early time data by eye but the cutoff values we used were
comparable to those that V. Kendon obtained by a more rigorous
procedure \cite{vivjfm}.) On the
other hand, it is found that when the domain size is much larger than
about a quarter of the simulation box, finite size effects start to be
relevant. Hence, from each simulation only a central portion of the
overall data can be relied upon; this data covers at most one decade
or so in time \cite{vivjfm}. As a result, the full crossover region
cannot be spanned by a single run but only by varying simulation
parameters to access $l(t)$ one section at a time. Since other
parameters, such as $\xi/L_0$, are varying during this procedure,
there is no guarantee, in general, of a smooth joining up of the
curves. In table \ref{table:parameters} we show the set of  fluid
parameters we have used to explore the crossover regime. We have 
also added a couple of sets that lie in the asymptotic regimes 
to check that the universal viscous and inertial scaling regimes 
are indeed recovered.

The initial period of diffusive growth, in each run, means that the
reduced physical time must be defined as $t = (T-T_{int})/T_0$, where
the offset time $T_{int}$ is not known a priori, and can vary from one
run to another. Previously, Kendon {\em et al.} developed a careful
procedure for fitting this parameter\cite{vivjfm}, separately for each
run, using the data from that run only. Although this is an objective
procedure, the fitted values of $T_{int}$ are very sensitive to the
details of the fitting, and the precise location of the resulting
$l(t)$ data is similarly sensitive. In the current work, therefore, we
leave $T_{int}$ as a free parameter for each run, and then globally
optimize the choices of all the $T_{int}$'s to achieve the best data
collapse on the $l(t)$ curve. This is best done by human eye using an
interactive graphics routine. The $T_{int}$ values are displayed in
table \ref{table:Tint}; given the uncertainties, these are broadly
consistent with the ones obtained previously.

\begin{figure}
\begin{center}
\includegraphics[width=12cm]{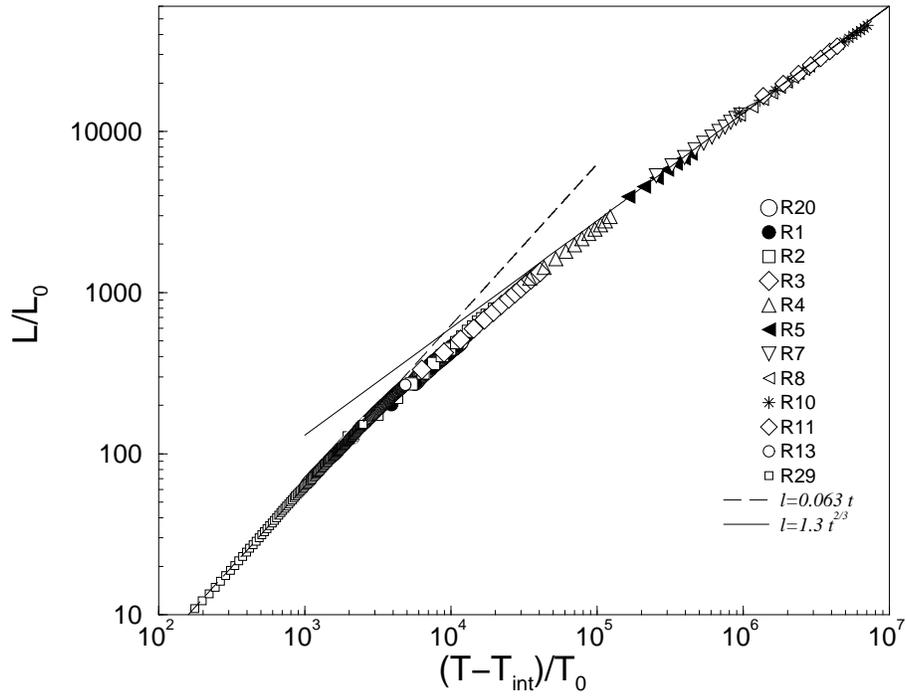}
\end{center}
\caption{Scaling plot in reduced variables for the runs of table \ref{table:parameters}.  Also shown the asymptotic theoretical predictions with amplitudes fitted from the simulations.}
\label{fig:scalgeneral}
\end{figure}

We show in Fig.\ref{fig:scalgeneral} the scaling curve we have
obtained in this fashion for all the runs of table
\ref{table:parameters}. One can clearly see that all the simulation
data can thereby be made to fall on a universal scaling curve, with
overlapping datasets extending from the viscous regime  deep into the 
inertial one. We have also drawn the two lines
that correspond to the viscous and inertial asymptotic laws to display
how these are attained at either end of the
crossover. Since we are using a definition for the domain size 
different from that of Ref.\cite{vivjfm}, we have to recompute the 
amplitudes of the asymptotic law. These are $b=0.065$ and $c=1.3$, 
which gives a crossover time $t^*=10^4$. Our crossover region now 
spans about two decades,
 somewhat narrower than the previous estimates; in
particular, runs that were previously taken to lie in the late
crossover \cite[Run29,Run30]{vivjfm,njp} are found to be much closer to the
asymptote than previously suggested. The reduced crossover width is
more in line with normal expectations and suggests that uncertainty in
the determination of $T_{int}$ by the previous method
\cite{vivprl,vivjfm} led to an exaggeration of the crossover.

\begin{figure}
\begin{center}
\includegraphics[width=12cm]{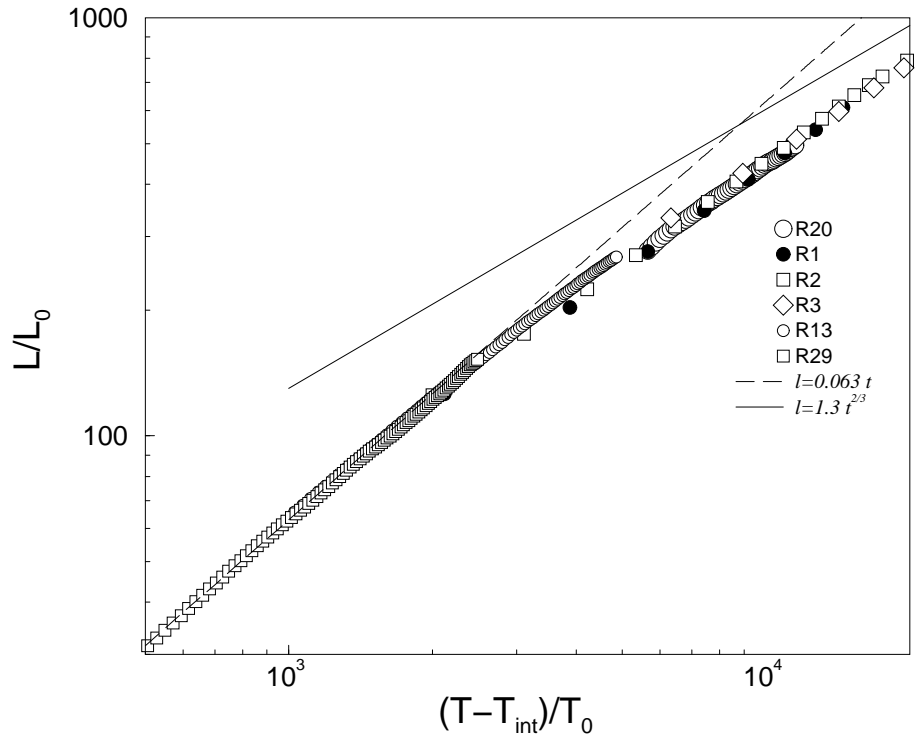}
\end{center}
\caption{Scaling plot in reduced variables for the runs of table \ref{table:parameters} that correspond to the transition from the viscous to the crossover regime.}
\label{fig:scalearly}
\end{figure}

In Fig.\ref{fig:scalearly} we zoom in on the initial
part of the crossover region. It is apparent how (to within small
errors),  the different runs
lie on top of each other, as pieces of a continuous universal curve.
It is interesting to look at Run13. Although in general, one run spans
only one dynamical regime (due to the reasons explained at the
beginning of the section), Run13 clearly exhibits a linear growth
followed by a deviation, at the beginning of the crossover region. That 
this is not a finite-size
artifact is clear, since runs that lie further in the crossover region
coincide with the final portion of Run13. 

\begin{figure}
\begin{center}
\includegraphics[width=12cm]{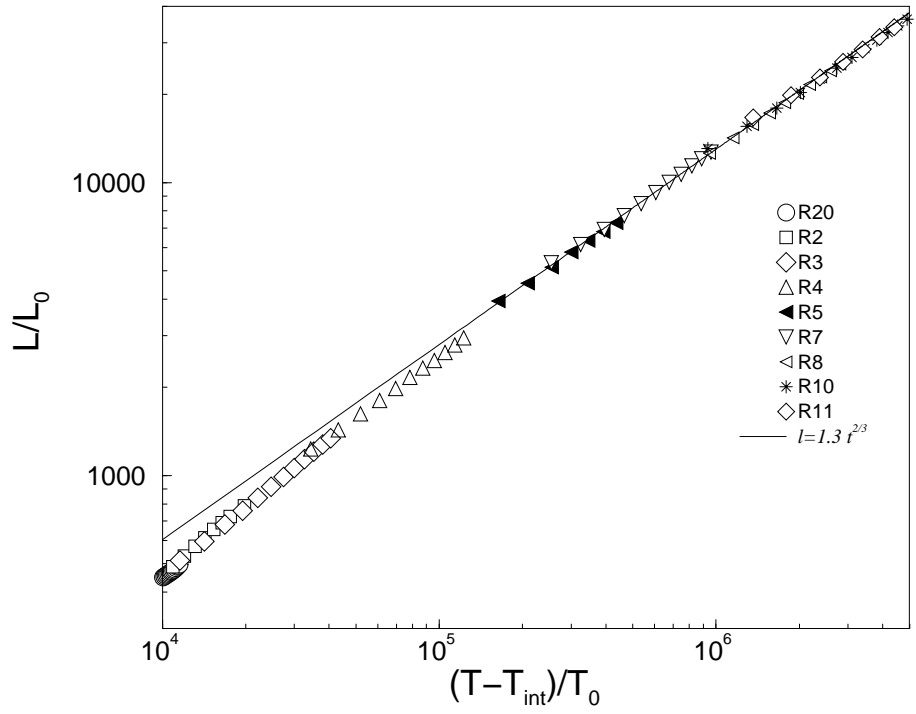}
\end{center}
\caption{Scaling plot in reduced variables for the runs of table
 \ref{table:parameters} that approach the inertial regime.}
\label{fig:scaldeep}
\end{figure}

In Fig.\ref{fig:scaldeep} we have focused on the late part of the
crossover region. Again, one can clearly see how different runs lie on
top of each other. The relaxation towards the final, inertial growth
law is faster than previously predicted. There is no evidence of any
kind of deviations. This nice collapse into a single curve rules out
the logarithmic deviations \cite{jury} that could be attributed to the
emergence of a new length scale in this dynamical regime.

Our procedure for choosing $T_{int}$ (by optimizing data collapse) is
less objective than the one used previously by Kendon et
al.\cite{vivjfm} and it is not surprising that the resulting scaling
is better. However, the sensitivity to
$T_{int}$, and the quality of the collapse when this is allowed to
float, leads us to conclude that there is no firm evidence {\em
against} dynamical scaling in the crossover region, despite the
previous results of Jury et al.\cite{jury}.

Of course, since the fitting is done by human eye, there is a danger,
perhaps, of seeing scaling where none exists. To check this, we have
tried the same procedure in two dimension where scaling is indeed held
not to exist (at least, not for bicontinuous
morphologies)\cite{wagner,wagner2D}. The same approach to the data
does not give similar collapse of the two dimensional data, which is
reassuring.

\begin{figure}
\begin{center}
\includegraphics[width=12cm]{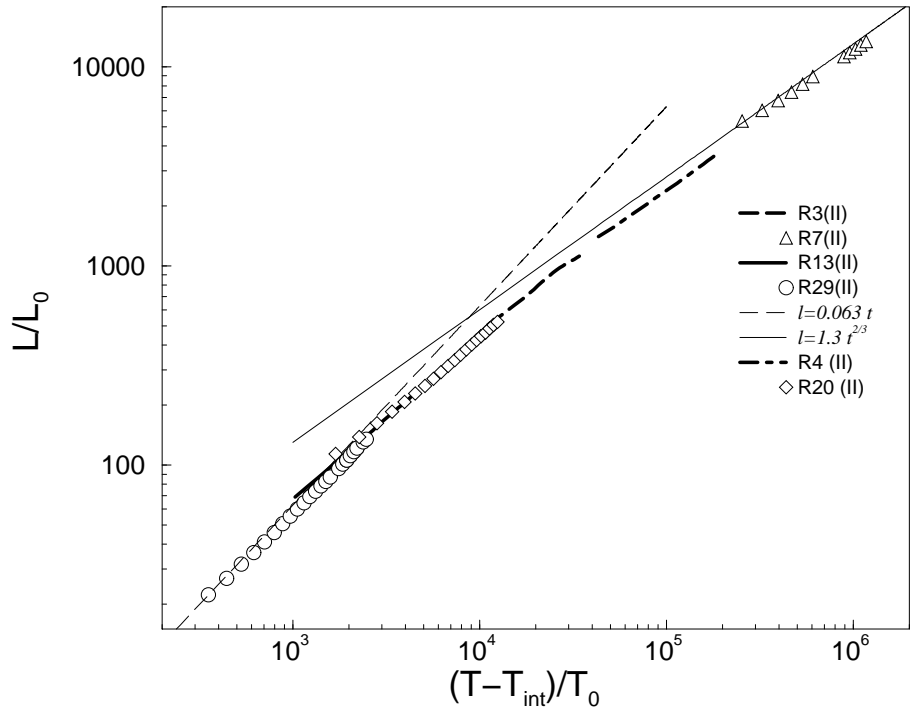}
\end{center}
\caption{Scaling plot in reduced variables for the runs of table
 \ref{table:parameters} computed using eq.(\ref{eq:lengthphi}).}
\label{fig:lphi}
\end{figure}

\begin{figure}
\begin{center}
\includegraphics[width=12cm]{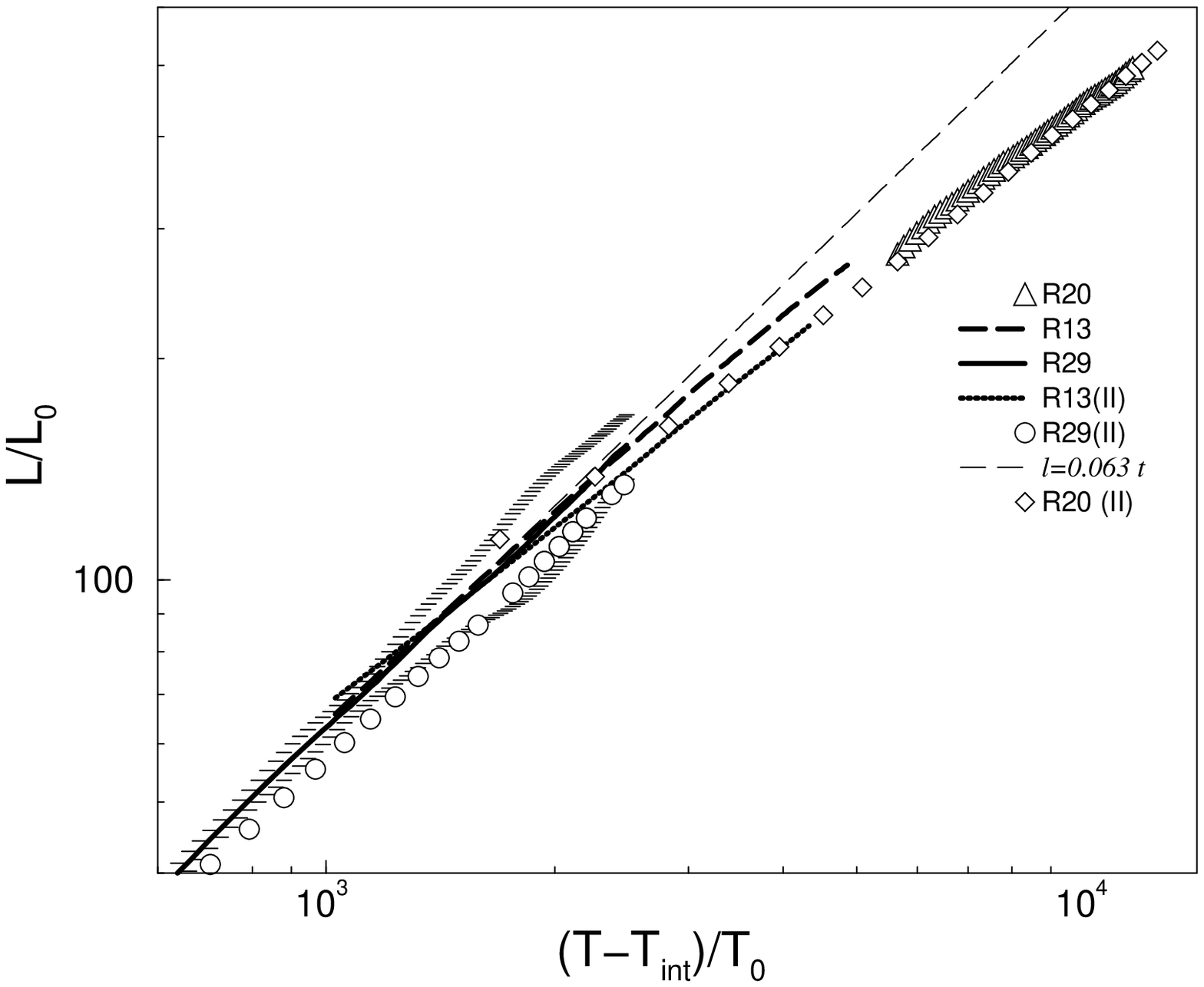}
\end{center}
\caption{Comparison of the scaled data computed using eqs.(\ref{eq:lt}) and
 (\ref{eq:lengthphi}). For run29 we show the estimated error bars in 
the determination of $L$ from eq.(\ref{eq:lt}) as upper and lower segments.}
\label{fig:compare}
\end{figure}

As a further
validation of our approach, we have also checked our results by
computing the domain size using an independent domain size
measurement. Specifically, for Runs 3, 4, 7, 13, 20 and 29 we have 
computed the spherically
averaged  order parameter structure factor \begin{equation} S(k,T)
\equiv \int_{|{\bf k}|=k}\phi({\bf k},T)\phi(-{\bf k},T) d^3{\bf k}
\end{equation} from it, we can obtain a measure of the mean domain
size as the inverse of its first moment, 
\begin{equation} L_{\phi}(T)
= \frac{\int S(k,T) dk}{\int k S(k,T) dk} \label{eq:lengthphi}
\end{equation} 
In this way we again sample the universal curve from the viscous up to
the inertial regime. We have used the values of $T_{int}$ obtained
from $L$ to construct the universal curve for $L_{\phi}$. In fig.
\ref{fig:lphi} we show the different runs after scaling appropriately,
which shows that we are able to collapse all the data on a single
universal curve. It exhibits the two asymptotic viscous and inertial
regimes.  The fact that we are able to use the obtained values of
$T_{int}$ to construct the universal curve using independent measures
of the domain lengths gives confidence that the scaling procedure we
have followed is robust. If we compare the $l(t)$ curves obtained from
the two domain lengths, we see that the ones derived from $L(T)$ tend
to lie above those corresponding to $L_{\phi}$.  However, in
fig.\ref{fig:compare} we compare the scaled curves obtained using the
two length measures for a few runs. For run 29 we have estimated the
error in the measure as the mean square deviation of three
length-scales measured as $1/\lambda_1$, $1/\lambda_2$ and
$1/\lambda_3$ (see eq. (\ref{eq:lt})), which we display as upper and lower
bars. The error in the determination of the domain size, which is
around $5\%$, does not allow to distinguish, in the late portion of
the runs, between the two measures. This analysis also shows that the
slight deviations from scaling visible in Fig. 2 are well within the error
tolerances expected\cite{note}.

\section{Conclusions}

In this paper we have analyzed in detail the crossover region of the
spinodal decomposition of a binary mixture. By performing a number of
new large lattice Boltzmann simulations, we have covered the regime
where viscous forces and inertial transport are simultaneously
relevant. Throughout the crossover, the data is compatible with the
existence of a universal scaling curve $l(t)$ in reduced physical
units. The crossover region is somewhat narrower
than previously reported. This can be attributed to the 
uncertainties in the behavior in the late part 
of the crossover arising from the ambiguities in the determination of
the fitting parameter $T_{int}$. 
The existence of the scaling curve argues
against the suggestion \cite{jury} that the physics of pinch-off
events can strongly interfere with the domain growth in this regime of
parameters.

\acknowledgements
This work was funded in part by EPSRC Grant GR/M56234 and EC Access to
Research Infrastructure Contract HPRI-1999-CT-00026 TRACS programme at EPCC.
.


\def\jour#1#2#3#4{{#1} {\bf #2}, #3 (#4).}
\def\tbp#1{{\em #1}, to be published.}
\def\tit#1#2#3#4#5{{#1} {\bf #2}, #3 (#4)}
\def\ap{Adv. Phys.}
\def\epl{Euro. Phys. Lett.}
\def\prl{Phys. Rev. Lett.}
\def\pr{Phys. Rev.}
\def\pra{Phys. Rev. A}
\def\prb{Phys. Rev. B}
\def\pre{Phys. Rev. E}
\def\pa{Physica A}
\def\ps{Physica Scripta}
\def\zpb{Z. Phys. B}
\def\jmpc{J. Mod. Phys. C}
\def\jpc{J. Phys. C}
\def\jpcs{J. Phys. Chem. Solids}
\def\jpco{J. Phys. Cond. Mat}
\def\jf{J. Fluids}
\def\jfm{J. Fluid Mech.}
\def\arf{Ann. Rev. Fluid Mech.}
\def\roy{Proc. Roy. Soc.}
\def\rmp{Rev. Mod. Phys.}
\def\jsp{J. Stat. Phys.}
\def\pla{Phys. Lett. A}

\renewcommand{\baselinestretch}{1}

\begin{table}
\begin{tabular}{|l|l|l|l|l|l|l|l|}\hline \hline
 Run number      & -A/B &   $\kappa$ &  $\eta$ & M & $\sigma$ & $L_0$
 & $t_0$ \\ \hline \hline
    29     & 0.0625  & 0.04   & 0.2   & 0.3  & 0.042   & 0.952 & 4.54\\
    13     & 0.00625 & 0.004  & 0.035 & 4.0  & 0.0042  & 0.29  & 2.43\\
    20     & 0.00625 & 0.004  & 0.025 & 4.0  & 0.0042  & 0.15  & 0.885\\
     1     & 0.03125 & 0.02   & 0.05  & 1.0  & 0.021   & 0.12  & 0.283\\ 
     2     & 0.00625 & 0.004  & 0.02  & 4.0  & 0.0042  & 0.095 & 0.45 \\
     3     & 0.00625 & 0.004  & 0.015 &  4.0 & 0.0042  & 0.054 & 0.19 \\ 
     4     & 0.00625 & 0.004  & 0.01  & 4.0  & 0.0042  & 0.024 & 0.0567 \\ 
     5     & 0.0125  & 0.008  & 0.0092& 2.0  & 0.0084  & 0.01  & 0.011 \\ 
     7     & 0.00625 & 0.004  & 0.005 & 4.0 & 0.0042  & 0.0059& 0.00709 \\ 
     8     & 0.00625 & 0.004  & 0.0035& 4.0  & 0.0042  & 0.0029& 0.00243 \\ 
     9     & 0.00313 & 0.002  & 0.00247&8.0  & 0.0021  & 0.0029& 0.00336 \\
    10     & 0.00313 & 0.002  & 0.00183&8.0  & 0.0021  & 0.0016&0.00139\\ 
    11     & 0.00625 & 0.004  & 0.0026 &4.0  & 0.0042  & 0.0016&  0.00099\\
\hline \hline
\end{tabular}
\caption{Parameters of the different runs and characteristic length
and times. Run20 corresponds to the notation of Ref.\cite{vivjfm}.}
\label{table:parameters} \end{table}

\begin{table}
\begin{tabular}{|l|c|c|c|c|c|c|c|c|c|c|c|c|}\hline \hline
 Run number &  29 & 13 & 20 & 1 & 2 & 3& 4 & 5 & 7 & 8 & 10 & 11 \\
 $T_{int}$  &  1200 & 700 & 1000 & 900 & 1200 & 1300 & 1550 & 1700 & 1700 & 1700 & 1700 & 1650\\
\hline \hline
\end{tabular}
\caption{ Computed values of the initial time $T_{int}$.}
\label{table:Tint} 
\end{table}

\begin{thebibliography}{2}
\bibitem{inertial}
In the latter, since the interfacial area and the kinetic energy are
both decreasing, the stated balance cannot be maintained in the energy
equation unless an energy cascade is introduced, in which the
reduction in interfacial area at large scales feeds kinetic energy
down to smaller scales where viscous dissipation finally removes
energy from the system. A three-length scales analysis predicts,
nonetheless, the same primary balance that can be obtained directly
from the Navier-Stokes equation (differences in the scaling of the
velocity fields are obtained, though)\cite{viv2prl}.
\bibitem{note} It is important to note that the initial noise
amplitude in the simulation is the same in all the runs, otherwise it
is not possible to compare the values of $T_{int}$, which depend on
the initial time neede to form the first domains.
\end{thebibliography}
\end{document}